\documentstyle[12pt]{article}
\begin{document}
%\hsize37truepc\vsize61truepc
%\hoffset=-.5truein\voffset=-0.8truein
%\setlength{\baselineskip}{17pt plus 1pt minus 1pt}
%This must be for that bizarre A4 paper they have in our Eastern Colonies
%\setlength{\textheight}{25.5cm}
%\vphantom{0}\vskip1.1truein
%
\title{\vspace{-1.in}The Phenomenology
of Strings and Clusters in the 3{--}$d$ Ising
Model}

\author{Vladimir~S.~Dotsenko\cite{landau}, Marco~Picco and Paul~Windey
\\[0.5em]
{\small LPTHE\cite{lpthe}}\\
{\small Universit\'e Pierre et Marie Curie, Bte 126, 4 Place Jussieu}\\
%{\small Bte 126, 4 Place Jussieu}\\
{\small 75252 Paris CEDEX 05, FRANCE}\\[0.5em]
Geoffrey~Harris and Enzo~Marinari\cite{rome}\\[0.5em]
{\small Physics Department and NPAC, Syracuse University}\\
%{\small Syracuse University}\\
{\small Syracuse, NY 13244, USA}\\[0.5em]
Emil~Martinec\\[0.5em]
{\small Enrico Fermi Institute and Department of Physics, University
of Chicago}\\
%{\small University of Chicago}\\
{\small Chicago, IL 60637, USA}\\[0.5em]}
\date{November 1993}
\maketitle
%\vfill
\begin{flushright}
  {\bf hep-th/9401129}\\
  {\bf SU-HEP-4241-563}\\
  {\bf PAR-LPTHE 93/59}

\end{flushright}
% ***********************************
%\newpage
\begin{abstract}
We examine the geometrical and topological properties
of surfaces surrounding clusters in the 3--$d$ Ising model.
For geometrical clusters at the percolation temperature and
Fortuin--Kasteleyn clusters at $T_c$, the number
of surfaces of genus $g$ and area $A$ behaves as $A^{x(g)}e^{-\mu(g)A}$,
with $x$ approximately linear in $g$ and $\mu$ constant.  We
observe that cross--sections of spin domain boundaries at $T_c$ decompose
into a distribution $N(l)$ of loops of length $l$ that scales
as $l^{-\tau}$ with $\tau \sim 2.2$.  We address the prospects
for a string--theoretic description of cluster boundaries.
\end{abstract}
\section{Introduction}

     One of the major successes of 20th century physics has been the
expression of the critical behavior of a variety of theories of nature
in terms of sums over
decorated, fluctuating paths.  It has thus also been hoped that
higher dimensional analogues, theories of fluctuating membranes, also
play a fundamental role in characterizing the physics of critical phenomena.
In particular, significant effort has been invested in
recasting one of the simpler models of phase transitions, the
3{--}$d$ Ising model \index{Ising model!in 3 dimensions},
as a theory of strings \cite{string} \index{String theory}.  These
attempts have
been stymied by the
difficulty in taking the continuum limit of formal sums over lattice
surfaces.

     In fact, sums over lattice surfaces, built from e.g.
plaquettes or polygons, generically fail to lead to a well-defined continuum
theory of surfaces.  An exception to this rule occurs when
the surface discretizations are embedded in $d \leq 1$.
In this case, one can exactly solve a large class of toy lattice models
which lead to sensible continuum `bosonic' string theories
(at least perturbatively) \cite{clt1}.  Numerically, it is observed that the
$d > 1$ versions of these
lattice models suffer a `fingering instability'; the embedded surfaces,
for instance are composed of spikes with thickness of the order of the
cutoff. It is suspected that the polygonal discretization of the worldsheet
(for large volumes)
is configured in a polymer{--}like structure, so that these theories
cannot be realized as sums over surfaces in the
continuum limit\index{Branched Polymers}.
This instability is anticipated theoretically, since the mass--squared
of the dressed identity operator of the bosonic string
becomes negative above $d=1$, presumably
generating an uncontrolled cascade of states that tear
the worldsheet apart\cite{tachyon}\index{Tachyon}.

    In the continuum limit, we know how to evade these problems
in special cases
through the implementation of supersymmetry and the GSO projection.  This
additional structure, however, leads to fundamental difficulties in
discretizing these theories.  In principle, one might hope to somehow
guess an appropriate continuum string theory and then show that it embodies
the critical behavior of a lattice theory, such as the 3--$d$ Ising
model.  The prospects for success through such an approach
seem rather poor at this time.

      Given this state of affairs, we have turned to a more
phenomenological approach, in which we attempt to generate `physical'
random surfaces in a particular model and then examine their
topological and geometrical properties.  We thus have chosen to look
at the structure of domain boundaries in the 3{--}$d$ Ising model.
The phenomenology of these self{--}avoiding
cluster boundaries is interesting in its own
right, since it describes a large universality class of behavior that is
expressed frequently and quite precisely by nature.  We
also might hope that our observations may be useful in gauging the
prospects of success of a string--theoretic description.  The
Ising model has been employed previously as a means to generate random
lattice surfaces \footnote{Through the use the phrase `lattice
surface' rather than `surface', we indicate that these objects should
{\it not} be necessarily inferred to be
real surfaces in the continuum limit.};
see for instance, the work of David \cite{david2}, Huse and Leibler
\cite{HuseLei}, Karowski and Thun
and Schrader \cite{previous}.  In a sense, this work extends these studies
by looking for new features of the geometry of these lattice surfaces; we also
consider boundaries of Fortuin--Kasteleyn clusters as well as
`geometrical' spin domains.  Much of our analysis consists of a measurement
of the distribution of surfaces as a function of their area $A$ and genus
$g$, $N_g(A)$
\footnote{The mean genus per Ising configuration is measured in references
\cite{previous}.  A determination of genus as a function of
area in an Ising system with anti--periodic boundary conditions have
also appeared recently \cite{Gliozzi}.}.   We shall determine the
functional form of $N_g(A)$.  We also perform block spin measurements
of the genus, to determine if a condensation of handles is present
on cluster boundaries at all scales.  These cluster boundaries
are strongly coupled and thus it appears cannot be directly characterized
by perturbative string theory.   We see that, however, boundaries of spin
domains at the Curie temperature are not just strongly--coupled
versions of the branched polymer--like objects that attempts to
build `bosonic' random surfaces typically generate.
They instead exhibit a richer
fractal structure,
albeit one not characteristic of surfaces.
We show that they obey a new scaling law
that describes the
distribution $N(l)$ of lengths $l$ of loops that compose cross--sections of
cluster boundaries.

\section{Ising Clusters and Surfaces}

      We shall begin by summarizing the basic physical properties
of the cluster boundaries that we have analyzed.
To a first approximation, a 2{--}dimensional membrane
of area $A$ and curvature matrix $K$
will exact an energy cost \cite{HuseLei,David}
\begin{equation}
\label{action}
H = \mu A + \lambda \int ({\rm{Tr}}K)^2 + \kappa \int {\rm{Det}}K;
\end{equation}
$\mu$ is the bare surface tension, $\lambda$ is referred to as the
bending rigidity and $\kappa$ couples to the Euler character of the
surface.  In the regime which characterizes random surfaces\index{Random
Surfaces},
the surface tension must be sufficiently small to allow
significant thermal fluctuations.
Note that the above action does not capture the entire dynamics;
it is essential also to keep in mind that
Ising cluster boundaries are naturally
self--avoiding.
We first consider
surfaces in the dual lattice that bound `geometrical clusters' formed
from sets of adjacent identical spins.  In this case, the Ising dynamics
generates an energy penalty proportional to the boundary area;
$\lambda_{bare}$ and $\kappa_{bare} = 0$.  The bare surface tension is
tuned by the Ising temperature.   To put this model in perspective,
we note that for real vesicles, for instance,
the couplings
$\lambda$ and $\kappa$ can be quite large; $\lambda$ ranging from
about $kT$ to $100kT$ have been measured \cite{David}.  The bending
rigidity may be irrelevant in the continuum limit, however.
%\footnote{One-loop renormalization group calculations indicate that this
%is so, though the results of numerical simulations are ambiguous.}.
The string coupling \footnote{We ignore
distinctions between intrinsic and extrinsic metrics.} is
equal to $\exp (-\kappa)$.
Through blocking spins, we make an estimate of the renormalization
group behavior of $\kappa$.
Unless $\kappa$ effectively becomes large in the infrared, the cluster
boundaries will fail to admit a surface description in the continuum limit.

       The geometrical clusters and their boundaries are {\it{not}}
present at all scales at the Curie
temperature\index{Clusters!Ising}.  Instead, for
temperatures somewhat below $T_c$ and all temperatures above $T_c$
two huge geometrical clusters comprise a finite fraction
of the entire lattice
volume.   These clusters percolate, that is, they wrap around
the entire lattice (we shall consider periodic boundary conditions).
Otherwise, the lattice only contains very small clusters that are the
size of a few lattice spacings; there are no intermediate size clusters.
We can understand this behavior by considering the $T \rightarrow
\infty$ behavior of these clusters.  Two percolated clusters span
the lattice even at infinite temperature, where clusters are smaller
than at $T_c$.  At $T = \infty$, the spins are distributed randomly
with spin up with probability $50\%$; the problem of constructing
clusters from these spins then reduces to pure site percolation with
$p=1/2$.  Pure site (or bond) percolation describes the properties of
clusters built by identifying adjacent colored bonds (sites), which are
colored randomly with probability $p$\index{Percolation}.  Above a critical
value $p=p_c$, the largest of these clusters percolates through
the lattice\cite{Stauffer}.
For the cubic lattice, it is known that an infinite cluster
will be generated (in the thermodynamic limit)
at $p_c \sim .311$. Thus, the fact that
the geometrical clusters have percolated in the high--temperature
regime and at the Curie point is
essentially a consequence of the connectivity of 3--$d$ lattices.

    At very low temperatures, however, there are few reversed (minority) spins
in the Ising model; these form a few small clusters.  As the density of
minority spins increases, the clusters become bigger until the largest
cluster percolates at some temperature $T_p < T_c$.
It has been suggested (see \cite{CambNaue} and
\cite{HuseLei})
that since this minority spin percolation appears to be due to an increase
in the concentration of minority spins and not to any long--distance
Ising dynamics, that this transition is in the same universality class
as pure (bond or site) percolation.  We emphasize that the scaling
of minority clusters
should not correspond to any non--analyticity in the thermodynamic
behavior of the Ising model; it should essentially be a `geometric effect'.

      There is another type of cluster, introduced by Fortuin and
Kasteleyn \cite{FK,CK}, that does proliferate over all length scales at the
Curie point\index{Clusters!Fortuin--Kasteleyn}.  These FK clusters
consist of sets of bonded spins; one draws these bonds between
adjacent
same--sign spins with a temperature dependent probability $p = 1 -
\exp(-2\beta)$.  Note that the geometrical clusters are built by
a similar procedure, using instead $p=1$.  FK clusters
arise naturally in the reformulation
of the Ising model as a percolating bond/spin model \cite{Sokal}.
For the Ising partition function can be recast as a sum over occupied and
unoccupied bonds with partition function
\begin{equation}
 Z = \sum_{bonds}p^{b}(1-p)^{(N_b - b)}2^{N_c}
\protect\label{BSPART}
\end{equation}
where $p = 1-\exp(-2\beta)$, $N_b$ denotes the number of bonds in
the entire lattice in which $b$ bonds are occupied and $N_c$ equals
the number of clusters that these occupied bonds form.  When the factor
$2^{N_c}$ is replaced by $q^{N_c}$, then (\ref{BSPART}) is the partition
function for the $q$-state Potts model.  If we
assign a spin to each bond so that all bonds in the same cluster
have the same spin, then the factor of $q^{N_c}$ just comes
from a sum over spin states.  The above partition function can then be viewed
as a sum over FK clusters.  Using this
construction, one can show that the spin-spin correlator in the original
Ising model is equal to the  pair connectedness function of FK clusters,
\begin{equation}
\langle \sigma(x)\sigma(y) \rangle = \langle \delta_{C_x,C_y}
\rangle ,
\protect\label{SSREL}
\end{equation}
which equals the probability that points $x$ and $y$ belong to
the same FK cluster \cite{Hu}.  It then follows that for $T \ge T_c$,
the mean volume of the FK clusters is proportional to the susceptibility
of the Ising model, so that indeed FK clusters only just start to
percolate at the Curie point.  Additionally, the relation (\ref{SSREL})
also implies that the spatial extent of the FK clusters is proportional to
the correlation length of the Ising model.
Furthermore, scaling arguments \cite{Wang}
demonstrate that at $T_c$, the volume distribution of FK clusters obeys
\begin{equation}
\label{vscaling}
N(V_{cl}) \simeq V_{cl}^{-\tau}, ~~~  \tau = 2 + \frac{1}{\delta},
\end{equation}
where $\delta$ denotes the magnetic exponent of the Ising model
($M \simeq B^{1/\delta}$).  Thus we see that FK clusters, unlike
the geometrical clusters previously discussed, directly
encode the critical properties of the Ising model.  Indeed, we are
necessarily led to study FK clusters in order to measure scaling laws
that characterize cluster boundaries of the scale of the Ising
correlation length, i.e. boundaries that scale at the Curie point.
On the other hand, geometrical cluster boundaries contribute an
energy penalty proportional to their individual area; the lattice
surface dynamics of FK cluster boundaries, however, cannot be likewise
described by a similar physical rule.

In 2--dimensions both the FK clusters and
the geometrical clusters
percolate at the Curie temperature.  The critical properties
of these clusters differ, however, since the scaling of geometrical clusters
is partially determined by the `percolative' properties of
two--dimensional lattices.  These effects are in some sense removed through
the FK construction.

\section{The Simulation}

We now proceed to outline the techniques used in our Monte Carlo
simulations\index{Monte--Carlo simulations}.
We analyzed data from lattices of size ranging from $L=32$ to $L=150$,
using about six months of time on RISC workstations.
Spin updates were implemented through the efficient Swendsen--Wang
algorithm \cite{SW}:  FK clusters for each lattice configuration
are first constructed, then the spins composing each cluster
are (all) assigned a new random spin value.
We determined our statistical uncertainties via the jackknife technique
and extracted exponents through linear least--squared fits.  Statistical
errors for these exponents were also obtained by using jackknife when
fitting.
Generally, systematic corrections to scaling and finite-size effects
are much larger than our statistical errors.

The main technical difficulty that we encountered was the measurement
of the Euler character, equal to $V - E + F$ for a dual surface with
$V$ vertices, $E$ edges and $F$ faces.  On the simple
cubic (SC) lattice, the construction of the dual surface
is ambiguous for configurations in which
4 plaquettes intersect along the same link, for instance.  We found that
we could define a consistent set of rules, which we shall discuss
in \cite{ourbigpaper}, that resolved these ambiguities.  These rules
are certainly not unique; one would hope that their implementation
essentially serves as a regularization that does not affect
long--distance scaling laws.

In two dimensions, one can avoid ambiguous intersections on the
dual lattice by considering Ising spins on the triangular lattice.
Its dual (the honeycomb lattice) is trivalent and thus Ising
spin domains will not be enclosed by
self--intersecting paths.  This fortuitous situation generalizes
to three--dimensions for
the Ising
model on a body centered cubic (BCC) lattice in which the vertices
at the center of each cube are also connected to those in the
centers of neighboring cubes.  More explicitly, we coupled with
equal strength both the $6$
nearest and $8$ next-nearest Ising spins so that only three plaquettes
of the dual lattice meet along a dual link.  Since surfaces built dual
to this lattice are also naturally self-avoiding, computing the genus
is trivial.  A depiction of the
Wigner--Seitz cell of this lattice (composed of plaquettes in the
dual lattice) appears
in figure 1.

Estimating the appropriate critical temperatures also required
considerable effort.  To find the percolation temperature $\beta_p$
we used the method discussed by Kirkpatrick \cite{Kirkpatrick} in
which one measures the fraction $f$ of configurations containing
clusters that span the lattice.  One plots $f$ versus $\beta$
for different lattice sizes $L$; $\beta_p$ corresponds to the
intersection of these curves for different $L$.   On the BCC
lattice, we checked this by also determining the temperature at which
the mean cluster size scales as a power law in $L$.  From this
analysis, we obtained $\beta_p = .0959$ on the BCC lattice and
$\beta_p = .232$ on the SC lattice.  The value of $\beta_c$
on the SC lattice has been previously determined to be
about $.221651$ \cite{Hasen}; we also found that $\beta _c \sim .0858$
on the BCC lattice.

\section{Results}

    We now present data from our simulations on both the
simple cubic and BCC lattices.  We have examined boundaries of
FK clusters at $T_c$, surfaces bounding minority spin domains
at $T_p$ and geometrical clusters at $T_c$.  In
\cite{ourbigpaper}, a more comprehensive discussion of our data
will appear, including also results from simulations of the 2--$d$
Ising model and pure bond percolation in 3 dimensions.  A more
concise summary of some of these results
has been presented in \cite{ourletter}.

\subsection{Cluster Properties}

We begin by discussing a few of the properties of the clusters.
Most of the new material, pertaining to the topology
of the cluster boundaries, appears in the following sub--sections.

    For FK clusters at $T_c$ and minority clusters at $T_p$,
we verified the scaling given in (\ref{vscaling}).  This is
shown, for example, in figure 2 for FK clusters
on an $L=64$ SC lattice.  We observe good power law scaling over a
large region of $V_{cl}$.  The bump in $N(V_{cl})$ and subsequent
steep drop--off describes the distribution of clusters that span
a significant fraction
of the lattice and are thus subject to large finite--size effects.
In principle, by carefully determining
$\tau = 2 + 1/\delta$, one might hope
to provide concrete numerical evidence for
the very reasonable hypothesis that the transition at $T_p$
is in the universality class of pure percolation.  In practice,
this is quite difficult. The value of the
magnetic exponent $\delta$ in the 3--$d$ Ising model (as determined through
renormalization group methods, for instance) yields the prediction
$\tau_{{\rm FK}} = 2.207(1)$ \cite{ItzyDrou}.  The value
of $\tau$ for pure percolation
that one would infer from recent series expansions is $\tau = 2.189 (5)$
\cite{Stauffer},
which is not so different from the FK value.

    In fact, the power law fits to $N(V_{cl})$
are not very precise, due to large finite volume
effects and corrections to scaling.
The values we extract from these
plots are $\tau_{{\rm FK}} =  2.25 (10)$
(this has been measured by Wang \cite{Wang})
and  $\tau_{{\rm geo}} = 2.10 (5)$
on the largest ($L=64$ and $100$) lattices that we considered.
This is a rather poor way to measure these exponents;
much more accurate estimates can be obtained through finite--size
scaling fits of the mean cluster size as a function of
lattice size $L$\index{Finite--size scaling}.   The mean
cluster size scales as $L^{\gamma/\nu}$;
standard scaling relations and (\ref{vscaling}) give
$\tau = (3 + \gamma/\nu d)/(1 +
\gamma/\nu d)$ ($d = 3$).  Using this technique , we
measured $\tau_{{\rm FK}} = 2.207(3)$
on the SC lattice
and $\tau_{{\rm geo}} = 2.202(3)$ on the BCC lattice.  The error
on $\tau_{{\rm geo}}$ is in fact probably several times larger than
quoted above, due to
uncertainties in locating the critical temperature.
This measurement of $\tau_{{\rm FK}}$ agrees
perfectly with previous values; the measurement of $\tau_{{\rm geo}}$
is not accurate enough to distinguish likely pure percolation
behavior from that of percolation of FK clusters.

We measured the number of sites on the boundary of each cluster, $A_{cl}$.
A plot of $\ln (V_{cl}/A_{cl})$ vs. $\ln (V_{cl})$ for FK clusters
on an $L=64$ BCC lattice
appears in figure 3.
We see that for very small volumes, the lattice
regularization constrains $V_{cl}$ to equal $A_{cl}$ and for intermediate
volumes, there is a small deviation from linear scaling (as some
interior sites begin to appear).  The plateau that appears around $V_{cl}
= 3000$ indicates the onset of scaling regime where $A_{cl} \propto V_{cl}$.
The growth just at the end of the plot is due to the largest cluster,
which wraps around the lattice and merges with itself to form extra
interior points.  This plateau indicates that the
lattice surfaces are not smooth and may be configured as
polymer--like networks.

This behavior is not surprising.
The observed proportionality of $V_{cl}$
and $A_{cl}$ is well-known in the context of pure percolation in $2$ and
$3$ dimensions \cite{Stauffer}.  Bonds (or sites) are deleted with
a fixed probability in percolation.  This implies that holes should
be distributed homogeneously with finite measure on percolation clusters;
i.e. the boundary length should be proportional to the enclosed volume.
Note that FK clusters
are constructed by performing percolation on geometrical clusters,
so this argument should definitely apply in the FK case.  We also found
$A_{cl} \propto V_{cl}$ for geometrical clusters at $T_p$; this observation
is consistent with the intuition that the $T_p$ transition
is that of pure percolation.

%The geometry of percolating clusters (for pure percolation) is
%described
%in more detail by the
%`links, nodes and blobs' picture
%in dimensions below $6$ \ref{Stauffer,DeGennes}.  The links form the
%thin backbones of the cluster; they are connected together at the nodes
%which occur roughly every correlation length $\xi$.  Most of the volume
%of the cluster consists of dangling ends emanating from the backbones.
%The backbones do not consist merely of one segment; they contain
%multiply-connected paths that form blobs with diameter
%up to size $\xi$.  Therefore, if this picture describes the transition
%at $T_p$ (and perhaps also schematically the behavior of percolating
%FK clusters), then the distribution of genus should essentially
%characterize the statistics of branches on the backbones.

\subsection{Genus Distribution}
We now turn to an analysis of the distribution of handles on
cluster boundaries.  If these boundaries form tangled networks, then
the following essentially characterizes the statistics of closed loops in
these networks.  In the simplest scenario, one might assume that
the handles are uncorrelated.  It would then follow
that $N_g(A)$ asymptotically obeys the Poisson distribution
$N_g(A)= \kappa_g (\mu A)^g e^{-\mu A}$, with $\kappa_g \propto 1/g!$.  The
probability per plaquette of growing a handle is then $\mu$.

We first present a sample of fits to $N_g(A)$ for
FK cluster boundaries on the BCC lattice for $L=64$.
In figure 4, we
present our data for genus $2$ along with a best fit to the functional
form
\begin{equation}
  \protect\label{funform}
  N_g(A) = C_g A^{x(g)} e^{-\mu (g)A}\ .
\end{equation}
%\epsffile{ng5_bcc.eps}
$N_2(A)$ is peaked near $A=250$, and the fit is perfect apart from the
very small area region, where we expect corrections to scaling to
be large.  Likewise, the power law plus exponential fit is superb
for genus $5$ as indicated in figure 5.
We find that this functional
form fits our data very well for $g \geq 2$ up to about $g = 20$ where
our statistics become poor.
If we assume that the ansatz (\ref{funform}) holds, then it follows that
\begin{equation}
    \mu = \mu_{eff} \equiv \frac{\langle A \rangle}{(\langle
  A^2 \rangle  - \langle A \rangle ^2) }
\end{equation}
and
\begin{equation}
   x(g) = x_{eff} \equiv \frac{\langle A \rangle ^2}{(\langle A^2
\rangle - \langle A \rangle ^2)} -1\ .
\end{equation}
We measured these moments and found that indeed $\mu_{eff}$ and $x_{eff}$
agreed very well with the values extracted directly from fits to
(\ref{funform}).
The value $\mu_{eff}^{-1}=114\pm 3$ as depicted in figure 6
is proportional to the average surface
area (in lattice units) per handle and is independent of genus
for $g>2$.
In figure 7 we show the genus dependence of the
exponent $x$, extracted both
from moments of the area distribution and from the direct fits.
After a transient region for small genus ($g=0-4$)
we find almost linear behavior in the region $g=5-15$ with a slope
of $1.25 \pm 0.1$.

The results for FK clusters on the SC lattice are quite similar, though
not as clean.  We first show the behavior of $N_1(A)$ for $L=64$ in
figure 8.
Clearly, here the fit does not work, though one does expect large
deviations from asymptotic scaling for surfaces in the range depicted.
The fit for genus $5$ (figure 9), however, is quite good, although small
systematic discrepancies are still notable.  Perhaps the regularization
needed to define genus in the SC case is partially responsible
for these deviations.
Again, for the SC lattice, we find that $\mu$ appears to
be independent of $g$, though we observe a very small systematic
drift.  The plot of $x(g)$ vs. $g$ exhibits more curvature
than in the BCC case, but the slope in the genus $5-15$ region again
is about $1.25$.

The deviation of $1.25$ from $1$ at first glance suggests
the presence of significant deviations from Poisson distributed
behavior.  This may be due, however, to systematic deviations
from continuum behavior due to lattice artifacts.
The magnitude of these systematic errors is illustrated by the
measurement of the dependence of the mean area on genus.
If we assume the ansatz (\ref{funform}) then we would predict that
the mean dual surface area $A$ should increase linearly with genus,
obeying
\begin{equation}
\protect\label{meanarea}
\langle A(g) \rangle = \frac{x(g) + 1}{\mu (g)}.
\end{equation}
% which
%can be measured much more
%accurately
% (since it does not depend on a fit or on a dispersion of
%moments).
For FK clusters on the BCC lattice, however,
we see by fitting ln$\langle A \rangle$ to ln$g$
in the small genus regime that $\langle A \rangle$ is not precisely linear
in $g$; in fact it scales roughly as $g^{.85}$.
Note that such a
scaling law could not hold asymptotically for large lattices and large
areas, since it would imply that surfaces would have more handles
than plaquettes.   Indeed this effective exponent slowly increases
with genus (to roughly $.90$ at $g=50$).   Thus we observe systematic
deviations (of order $15\%$) of genus dependent exponents
from their asymptotic
values.
This also indicates that the apparent linearity that we observed in
$x(g)$ is somewhat deceiving; presumably deviations from linearity
would be more apparent if our statistics were better and we
could directly fit somewhat higher values of $g$.
The slope of $x(g)$ should decrease with greater $g$, so that
the above estimate of the slope ($1.25$) may be too large.

   The genus behavior of geometrical clusters at $T_p$ is qualitatively quite
similar to that of the FK case just discussed.  We show fits to $N_2(A)$ and
$N_5(A)$ on an $L=60$ lattice in figures 10 and 11.
There are large deviations in the fit for genus $2$; for genus $4$
and larger, however, the fits are nearly perfect.  Again, $\mu$
is approximately independent of $g$, though (from figure 12)
we observe transient
behavior that is very significant up to genus 10.
Again, $x$ is approximately linear in $g$ (as shown in figure 13), with
a slope considerably
lower than in the FK case;
$dx/dg\sim 0.7\pm 0.1$ in the range $g=3-40$ for Ising
minority spin percolation.  The same caution as before applies
to these
slope values; systematic errors could still be quite large,
so the actual value of $.7$ for the slope is not so trustworthy.
In this case, potential deviations from
asymptotic scaling reveal themselves
most clearly through the transient behavior of $\mu$.

 From this analysis, we can conclude that the genus distribution
of FK cluster boundaries at $T_c$ and geometrical cluster boundaries
at $T_p$ is described by the functional form
(\ref{funform}),
with $x(g)$ approximately linear in $g$ and $\mu$ constant.
The lattices considered, however, are too small to characterize
the behavior of $x$ more precisely.

\subsection{Loop Scaling and Blocked Spins}

  In this section, we will solely be concerned with the structure
of boundaries of geometrical clusters as $T$ is increased beyond $T_p$,
particularly to $T=T_c$.  Recall that for $T > T_p$, two percolated
clusters of opposite sign will span the lattice.  For $T$ not so close
to $T_c$, we expect that the characteristics of the Ising interaction
will not influence the large--scale structure of these percolating clusters.
The percolating clusters (assuming the transition at $T_p$ is indeed
in the universality class of pure percolation) should then be described
by the `links, nodes and blobs' picture developed for the infinite
clusters of pure percolation in dimensions below $d_c=6$
\cite{Stauffer,DeGennes}\index{Links, nodes and blobs}.
In this description, the links form the
thin backbones of the cluster; they are connected together at the nodes
which occur roughly every percolation correlation
length $\xi$.  Most of the volume
of the cluster consists of dangling ends emanating from the backbones.
The backbones do not consist merely of one segment; they contain
multiply-connected paths (which close to form the handles
that we measure) that form blobs with diameter
up to size $\xi$.

     A cross section of the boundaries of these networks
of tangled thin tubes would presumably be composed of a set of
small lattice--sized loops.  To check this, we examined the phase
boundaries between up and down spins on planar slices of both
the SC and BCC lattices.  In figure 14, we show a
log--log plot of $N(l)$, the number of loops of length $l$, versus
$l$ taken at the percolation temperature $
\beta_p = .232$ on the SC lattice.  The
curve exhibits a sharp drop--off, indicating indeed
that these slices contain only small loops.
As we dial the temperature up towards $T_c$, we find that
larger loops begin to appear in the slices.  In fact,
at $T_c$, we find loops at all scales; $N(l) \sim l^{-\tau '}$!
This scaling is depicted in the log--log plot in
figure 15.  As in figure 2, we observe a small bump at the end
of the distribution followed by a rapid drop--off.  These deviations
from scaling are again due to the influence of the finite--size
of the lattice on the largest loops.
All of the largest loops must bound the
two percolating clusters, since there are no intermediate size
geometrical clusters at $T_c$.
The loops themselves have a non--trivial fractal structure;
we determined that the number of sites enclosed within a loop of length $l$
scales as $A(l) \sim l^{\delta '}$.

 From these measurements, we estimated that $\tau ' = 2.06(3)$ and
$\delta ' = 1.20(1)$.  These values are probably not very accurate,
however.  As in the determination of $\tau$ from the
behavior of $N(V_{cl})$, corrections to scaling and finite--size effects
are a source of large systematic errors.  These systematic
effects were only of order $1-2\%$ for $\delta$; thus we suspect that
our estimate of $\delta '$ is considerably better
than that of $\tau '$.  Carrying out these measurements also
required a resolution of certain ambiguities.  In particular, since
the boundaries of domains
self--intersect on slices of the cubic lattice, we had
to pick a prescription (effectively another short--distance regularization)
to define loops.  Additionally, the enclosed area is not
well--defined for loops
that wind around the (periodic) lattice.  We thus chose
to exclude loops with non--zero winding number from
consideration.  Also, we note that these measured values presumably
suffer from large systematic corrections because they do not
satisfy the
relation $\tau ' = 1 + \delta '$, which
can be derived through scaling arguments \footnote{We thank
Bertrand Duplantier \cite{priv} for providing
us with a derivation of this relation.}.
This relation also holds for the corresponding indices that
describe the distribution of self--avoiding loops that bound
clusters in the 2--$d$ Ising model at the Curie temperature.  In that
case, $\tau ' \sim 2.45$. Finally, we found that the scaling
behavior of loops on slices
slowly disappeared as we
continued to increase the Ising temperature.  At
$\beta = .18$ on $L=150$ SC lattices, we observed that very large loops
were again exponentially suppressed in the distribution $N(l)$.

     Should we be surprised by the presence of this `loop scaling' at
$T_c$?  The following argument, due to Antonio Coniglio, indicates that
this result is at least plausible \cite{priv2}.  First, note that in the  $T
\rightarrow \infty$ limit, the distribution of loops and geometrical
clusters is that of pure site percolation with $p = .5$.  On the
square lattice, $p_c \sim .59$ so that if only half the sites
contain identical spins, then the distribution of loops
and clusters should be
governed by a finite correlation length. Now consider turning on
the Ising couplings in the $x$ and $y$ directions.  As the spins
become correlated, the critical concentration\footnote{Note
that we can adjust the relative concentration of up and down spins
by also adding a magnetic field.} needed for percolation should
diminish.  At the Curie temperature for the 2--$d$ Ising model
($T_c^{d=2}$) this critical concentration decreases to $.5$
and geometrical clusters and their boundaries percolate.
In two dimensions, this critical concentration cannot be less
than $.5$, since generically two percolating clusters cannot
span a single lattice \cite{Conigliopap}.   Imagine next turning
on the Ising coupling in the $z$ direction while tuning
the $x$ and $y$ couplings to remain at criticality.  If
the critical concentration remains $.5$ as the system reaches
the 3--$d$ Curie temperature, then one would find a scaling
distribution of clusters and boundaries on 2--$d$ slices.
On the other hand, we cannot rule out the possibility that the
critical concentration again increases above $.5$; then
we would never expect to find scaling of loops on slices of
the 3--$d$ Ising model.
%  Of course, we cannot dismiss
%the unlikely possibility that the loops on 2--$d$ slices
%are cut off in size at $T_c$ at a length scale well beyond
%the lattice sizes used in our simulations.

      We also observed scaling behavior of loops on the BCC lattice.
In particular, only small loops were found at $T_p$ while scaling
of $N(l)$
with the values $\tau ' = 2.23(1)$ and $\delta ' = 1.23(1)$
occurred at $T_c$.  The uncertainty in the value of $T_c$ probably
leads to a significant systematic error in the estimate of these
exponents.  They do obey the anticipated relation $\tau ' = 1 +
\delta '$; $\delta '$ is not particularly far from the estimate
extracted from the SC data.  Note that on slices of the BCC lattice, which
are triangular, there is no longer any ambiguity in the definition
of loops.
In this case, $N(l)$ apparently satisfies a power--law
distribution, with a temperature--dependent exponent, for all $T>T_c$!
This observation can be fully understood theoretically, since
the percolation threshold
on triangulated lattices equals $.5$.  Therefore, we definitely
expect to observe loop scaling at $T =
\infty$ with scaling exponents characteristic of 2--$d$ percolation
($\tau ' \sim 2.05$ and $\delta ' = 1$).  Since lowering the temperature
increases correlations between spins, we expect to find percolated
clusters on slices for all $T$.  For $T<T_c$, however, minority
spins cannot percolate on 2--$d$ slices because, as stated above,
only one infinite cluster can span a lattice.  Thus the minority
spins and the loops that enclose them must percolate at $T_c$
on 2--$d$ slices of the 3--$d$ Ising
model on the BCC lattice.  If we assume that this phenomenon is
independent of the particular lattice type, then it follows that
loop scaling should always occur at $T_c$.  A similar situation occurs
for the 2--$d$ Ising model on the triangular lattice:  one can
argue that the distribution
$N(l)$ again scales as a power law for all $T > T_c$ because $p_c = 1/2$
on triangulated lattices.

It also seems reasonable that the presence of loop scaling
may be related to the vanishing of the surface tension of
the Ising model at $T_c$.  The vanishing of
the surface tension ensures that the free energy
of a system with anti--periodic boundary conditions along one plane
(essentially due
the insertion of a large loop along the boundary) equals the
free energy of a system with periodic boundary conditions\index{Ising
model!surface
tension}.

We now comment on the significance of this scaling.
  As we noted in the previous two sub--sections, the geometrical
cluster boundaries do not in the least resemble surfaces (in
the continuum limit) at $T_p$.  The presence of large loops at
$T_c$ might indicate that the boundaries grow large long handles.
A visual examination of successive slices qualitatively indicates
that this is not so.  Large loops seemingly always vanish after
several consecutive slices.  Indeed, it is difficult to envision
a smooth surface that decomposes into a scaling distribution of loops
along arbitrary slices.

  It should also be noted that the exponent $\tau '$
is probably not directly related to the magnetic or thermal exponents
of the 3--$d$ Ising model.  More generally, it may not
be associated with the behavior
of correlation functions of local operators in a unitary
quantum field theory.
This is true also for loops bounding
clusters in the 2--$d$ Ising model.  For in all of these cases,
the scaling of geometrical clusters
is determined by the geometric effects associated with percolation as
well as the long--range correlations due to Ising criticality.
Still, this scaling law describes physics that in
principle is observable, perhaps by counting domains in sections of
crystals that lie in the universality class of 3--$d$ Ising.
It would thus be quite interesting to construct a theoretical
scheme to compute (approximately) the value of $\tau '$.
These loops are significantly `rougher' than the corresponding
boundaries in the 2--$d$ Ising model, since the exponent $
\delta ' $ is lower here.  They gain more kinetic
energy because they are given an extra dimension in which to
vibrate; perhaps this is responsible for their increased roughness.

Ideally, we would like
view these loops as string states that evolve in Euclidean
time (perpendicular to the slices).  Their dynamics
is described by the transfer matrix determined from Boltzmann
factors associated with their creation, destruction, merging and
splitting\index{Transfer matrix}.
We have thus found that the ground state wave functional (string field)
of this transfer matrix is peaked around
configurations that describe a scaling distribution of loops.
These loops seemingly bear little relation to free strings,
though, because they interact strongly by splitting and joining
every few lattice spacings \footnote{In practice, this makes an
analysis of the transfer matrix a formidable task.}.  One
might hope that some sort of
perturbative string
description could still be viable if the strength of this interaction
were just a short--distance artifact; i.e. if the string coupling
diminished towards zero in the infrared.  To gauge whether this
is likely, we blocked spins in our simulations to measure the renormalization
group flow of the operator that couples to the total Euler character
summed over all cluster boundaries.  In particular, during simulations
on L=128 SC and BCC lattices, we blocked spins, using the majority
rule and letting our random number generator decide
ties\index{Renormalization group!Blocked spins}.  At
each blocking level, we reconstructed clusters and boundaries and
then measured the genus summed over surfaces.  We present the results
of this analysis in table \ref{bstable}; data was taken at $\beta_c =
.221651$ on the SC lattice and $\beta_c = .0858$ on the BCC lattice.
\begin{table}
\begin{center}
\begin{tabular}{|l||l|l|l|l|l|} \hline
 lattice & 128 & 64 & 32 & 16 & 8 \\
\hline
BCC & .049 (3) & .039 (3) & .037 (3) & .039 (3) & .044 (3) \\
\hline
SC  & .021 (2) & .020 (2) & .018 (2) & .015 (2) & .012 (1) \\
\hline
\end{tabular}
\end{center}
\protect\caption{\label{bstable}The mean genus per lattice site
at $T_c$ for blockings ($L=8,16,32$ and $64$) of an $L=128$ lattice.}
\end{table}

The results are not so conclusive.  In particular,
since we lack a very precise determination of the Curie temperature on the
BCC lattice, it is likely that by the final blocking the couplings have
flowed significantly into either the high or low--temperature regimes.
Thus, one should probably not take the increase in genus density
in the final two blockings on the BCC lattice seriously.  This effect
is not a problem on the SC lattice, where we fortunately know the
critical temperature (based on previous Monte Carlo Renormalization
Group measurements) to very high accuracy.  On the other hand, we
suspect that the small $L$ blocked values on the SC lattice may be
unreliable, due to ambiguity in the definition of genus.  We can at
least infer that the genus density decreases a bit during the first
few blockings, indicating that the coupling $\exp (-\kappa)$
does at least slowly diminish at the beginning of the RG flow.  There is
no clear indication, however, that the flow continues on to the weak
string coupling regime.
One might also object to our choice of blocking scheme.
Indeed, perhaps it might be more appropriate to somehow block
the cluster boundaries themselves rather than the spins.
In practice this
would probably be technically difficult.

\section{Assessment}

     The prospects for passing from the Curie point to the regime
in which surfaces are weakly coupled are addressed in the work of
Huse and Leibler \cite{HuseLei}.  They qualitatively
map out the phase diagram of
a model of self--avoiding surfaces with action (\ref{action}).
The large $\kappa$ (large coupling to total Euler character) regime
of their model lies in a droplet crystal phase, where the large
percolated surface has shattered into a lattice of small disconnected
spheres.  Such a configuration maximizes the Euler density; it
clearly does not correspond to a theory of surfaces.  By
estimating the free energy difference between phases, they argue
that the transition to this droplet crystal is first order.  Given
this picture, there seems to be little evidence for the existence of
a fixed point describing a weakly coupled theory of surfaces
near the Curie point of the Ising model.  Nevertheless, we cannot
definitely exclude the possibility that there is still some path
which we have not considered to a weak--coupling theory.

 In conclusion, it appears that evidence of a continuum theory
of surfaces has eluded us in our investigation of Ising cluster
boundaries.   We have found, however, that these cluster boundaries
do exhibit an intriguing fractal structure that does not typically
appear in models of lattice surfaces.

\section{Acknowledgments}
We would like to thank Stephen Shenker for essential discussions which
led to our investigations.  We also greatly benefitted from
discussions and correspondence with Mark Bowick, Antonio Coniglio,
Francois David, Bertrand Duplantier,
John Marko and Jim Sethna.  Furthermore, we are indebted to
the organizers of this workshop for their efforts and hospitality.
We are also grateful to NPAC for their
crucial support.
This work was supported in part by the Dept. of Energy grants
DEFG02-90ER-40560, DEFG02-85ER-40231, the Mathematical Disciplines
Institute of the Univ. of Chicago, funds from Syracuse Univ., by the
Centre National de la Recherche Scientifique, by INFN and the EC
Science grant SC1*0394.

%\section{Index}
\index{Branched polymers}
\index{Ising model!in 2 dimensions}
\index{Ising model!in 3 dimensions}
\index{Ising model!surface tension}
\index{Percolation}
\index{Renormalization Group!blocked spins}
\index{Clusters!Ising}
\index{Clusters!Fortuin--Kasteleyn}
\index{Finite--size scaling}
\index{Links, nodes and blobs}
\index{Monte--Carlo simulations}
\index{String theory}
\index{Tachyon}
\index{Transfer matrix}

\vfill
\newpage
\topmargin -.3in
\flushbottom
\samepage{
\begin {flushleft}
{\bf Figure Captions}
\end{flushleft}
\begin{itemize}
\item[Fig.~1] The Wigner--Seitz cell of the BCC lattice with next--nearest
neighbor interactions.

\item[Fig.~2] $\ln N(V_{cl})$ vs. $\ln V_{cl}$ for FK clusters
on a $N=64$ SC lattice.

\item[Fig.~3] ln($V_{cl}$/$A_{cl}$) vs. ln$V_{cl}$ for FK clusters on the
$L=64$ SC lattice.

\item[Fig.~4] The number of genus $2$ surfaces at $T_c$
as a function of dual
surface area $A$ for FK clusters on the $L=64$ BCC lattice, with a best
fit to the functional form given in equation \ref{funform}.

\item[Fig.~5] As in the previous figure, but for genus 5.

\item[Fig.~6] The dependence of  $\mu$ (extracted from the
moments of the area distribution) on genus for FK clusters
on the $L=64$ BCC lattice at $T_c$.

\item[Fig.~7] The dependence of $x$ (extracted from
direct fits to (\ref{funform})
and moments) on genus for FK clusters on the $L=64$ BCC lattice at $T_c$.

\item[Fig.~8] The number of genus $1$ surfaces
at $T_c$ as a function of dual
surface area $A$ for FK clusters on the $L=64$ SC lattice, with
a best fit to the functional form (\ref{funform}).

\item[Fig.~9] As in the previous figure, but for genus 5.

\item[Fig.~10] The number of genus $2$ surfaces at $T_p$ as a function of
dual surface area $A$ bounding minority (geometrical) clusters on the
$L=60$ BCC lattice.

\item[Fig.~11] As in the previous figure, but for genus 5.

\item[Fig.~12] The dependence of $\mu$ (extracted
from moments) on genus for
surfaces bounding minority (geometrical) clusters on the $L=60$ BCC
lattice at $T_p$.

\item[Fig.~13] The dependence of $x$ (extracted from fits
and moments) on genus
for surfaces bounding minority (geometrical) clusters on the $L=60$ BCC
lattice at $T_p$.

\item[Fig.~14] A log--log plot of the distribution of loops
of length $l$ on slices of an $L=60$ SC lattice at $T_p$.
\item[Fig.~15] A log--log plot of the distribution of loops of
length $l$ on slices of an $L=150$ SC lattice at $T_c$.
\end{itemize}
}

\newpage
\begin{thebibliography}{99}
  \bibitem[*]{landau}
    Also at the Landau Institute for Theoretical Physics, Moscow
  \bibitem[\dag]{lpthe}
    Laboratoire associ\'e No. 280 au CNRS.
  \bibitem[\ddag]{rome}
    Also at Dipartimento di Fisica and INFN, Universit\`a di Roma Tor
    Vergata, Viale della Ricerca Scientifica, 00133 Roma, Italy.
%\bibitem{Regge} T. Regge, Nuovo Cimento 19 (1961) 558.
  \bibitem{string}
    E. Fradkin, M. Srednicki and L. Susskind, Phys. Rev. {\bf D21},
    (1980) 2885;
    C. Itzykson, Nucl. Phys. {\bf B210} (1982) 477;
    A. Casher, D. F\oe rster and P. Windey, Nucl.
    Phys. {\bf B251} (1985) 29;
    Vl. Dotsenko and A. Polyakov, {\it in} Advanced Studies in Pure
    Math. {\bf 15} (1987).
  \bibitem{clt1}
   E. Br\'{e}zin and V.A. Kazakov,
    Phys. Lett. {\bf 236B} (1990) 144;
    M.R. Douglas and S.H. Shenker,
    Nucl. Phys. {\bf B335} (1990) 635;
    D. J. Gross and A. A. Migdal,
    Phys. Rev. Lett. {\bf 64} (1990) 127.
  \bibitem{tachyon}
   G. Parisi,
   {\it in} {\em Proceedings of the Third Workshop on
   Current Problems in High Energy Particle Physics},
    John Hopkins Conference, Florence 1979;
    G.~Parisi,  J.-M.~Drouffe and N.~Sourlas, Nucl. Phys.
    {\bf B161} (1979) 397;
    B.~Durhuus, J.~Frohlich and T.~Jonsson,  Nucl. Phys. {\bf B240} (1984)
       453;
    J.~Ambjorn, B.~Durhuus, J.~Frohlich and P.~Orland,  Nucl. Phys.
    {\bf B270} (1986) 457;
    M.~E.~Cates, Europhys. Lett. {\bf 8} (1988) 719.
%  \bibitem{solvers}
%    See for example
%    O. Alvarez, E. Marinari and P. Windey, {\em Random Surfaces and
%    Quantum Gravity,} (Plenum, New York USA 1992) and references therein;
%    F. David, {\em Simplicial Quantum Gravity and Random Lattices,}
%    to be published;
%
%    P. Ginsparg and G. Moore, {\em Lectures on $2d$ Gravity and $2d$
%    String Theory,} to be published.
%  \bibitem{kpz}
%    K.~G.~Knizhnik, A.~M.~Polyakov and A.~B.~Zamolodchikov,
%    Mod. Phys. Lett. {\bf A3}, 819 (1988);
%    F.~David, Mod. Phys. Lett. {\bf A3}, 1651 (1988);
%    J.~Distler and H.~Kawai, Nucl. Phys. {\bf B321}, 509 (1989).
  \bibitem{david2}
    F.~David, Europhys. Lett. {\bf 9} (1989) 575.
  \bibitem{HuseLei}
    D.~Huse and S.~Leibler, J. de Physique {\bf 49} (1988) 605.
  \bibitem{previous}
    M. Karowski and H.J. Thun, Phys. Rev. Lett. {\bf 54} (1985) 2556;
    R. Schrader, J. Stat. Phys. {\bf 40} (1985) 533.
  \bibitem{Gliozzi}
    M. Caselle, F. Gliozzi and S. Vinti, Turin Univ. preprint DFTT--12--93.
  \bibitem{David}
    F. David, Jerusalem Gravity (1990) 80.
  \bibitem{Stauffer}
    D.~Stauffer and A.~Aharony,
    {\em Introduction to Percolation Theory,}
    (Taylor and Francis, London, U.K. 1992).
  \bibitem{CambNaue}
    J. Cambier and M. Nauenberg, Phys. Rev. {\bf B34} (1986) 8071.
  \bibitem{FK}
    C.M. Fortuin and P.W. Kasteleyn, Physica {\bf 57} (1972) 536.
  \bibitem{CK}
    A.~Coniglio and W.~Klein, J. Phys. {\bf A13} (1980) 2775.
  \bibitem{Sokal}
    R.G.~Edwards and A.D.~Sokal, Phys. Rev. {\bf D38} (1988) 2009.
  \bibitem{Hu}
    C.-K. Hu, Phys. Rev. {\bf B29} (1984) 5103.
  \bibitem{Wang}
    J.-S. Wang, Physica {\bf A161} (1989) 149.
  \bibitem{SW}
    R.H. Swendsen and J.-S. Wang, Phys. Rev. Lett. {\bf 58} (1987) 86.
  \bibitem{ourbigpaper}
    V.~Dotsenko, G.~Harris, E.~Marinari, E.~Martinec, M.~Picco and P.~Windey,
    in preparation.
  \bibitem{Kirkpatrick}
    S.  Kirkpatrick, in ``Ill-Condensed Matter'',  Les Houches Proceedings,
    Vol.31, ed. R. Balian, R. Maynard and G. Toulouse
    (North Holland, Amsterdam 1983) 372.
  \bibitem{Hasen}
    M. Hasenbusch and K. Pinn, Munster Univ. preprint MS--TIP--92--24.
  \bibitem{ourletter}
    V.~Dotsenko, G.~Harris, E.~Marinari, E.~Martinec, M.~Picco and P.~Windey,
    Phys. Rev. Lett. {\bf 71} (1993) 811.
  \bibitem{ItzyDrou}
    C.~Itzykson and J-M. Drouffe, {\em Statistical Field Theory},
    Cambridge University Press, Cambridge (1989).
  \bibitem{DeGennes}
    P.G. DeGennes, La Recherche {\bf 7} (1976) 919.
  \bibitem{priv}
    Bertrand Duplantier, private communication.
  \bibitem{priv2}
    Antonio Coniglio, private communication.
  \bibitem{Conigliopap}
    A. Coniglio, C. Nappi, F. Peruggi and L. Russo, J. Phys. {\bf A10}
     (1977) 205.
\end{thebibliography}
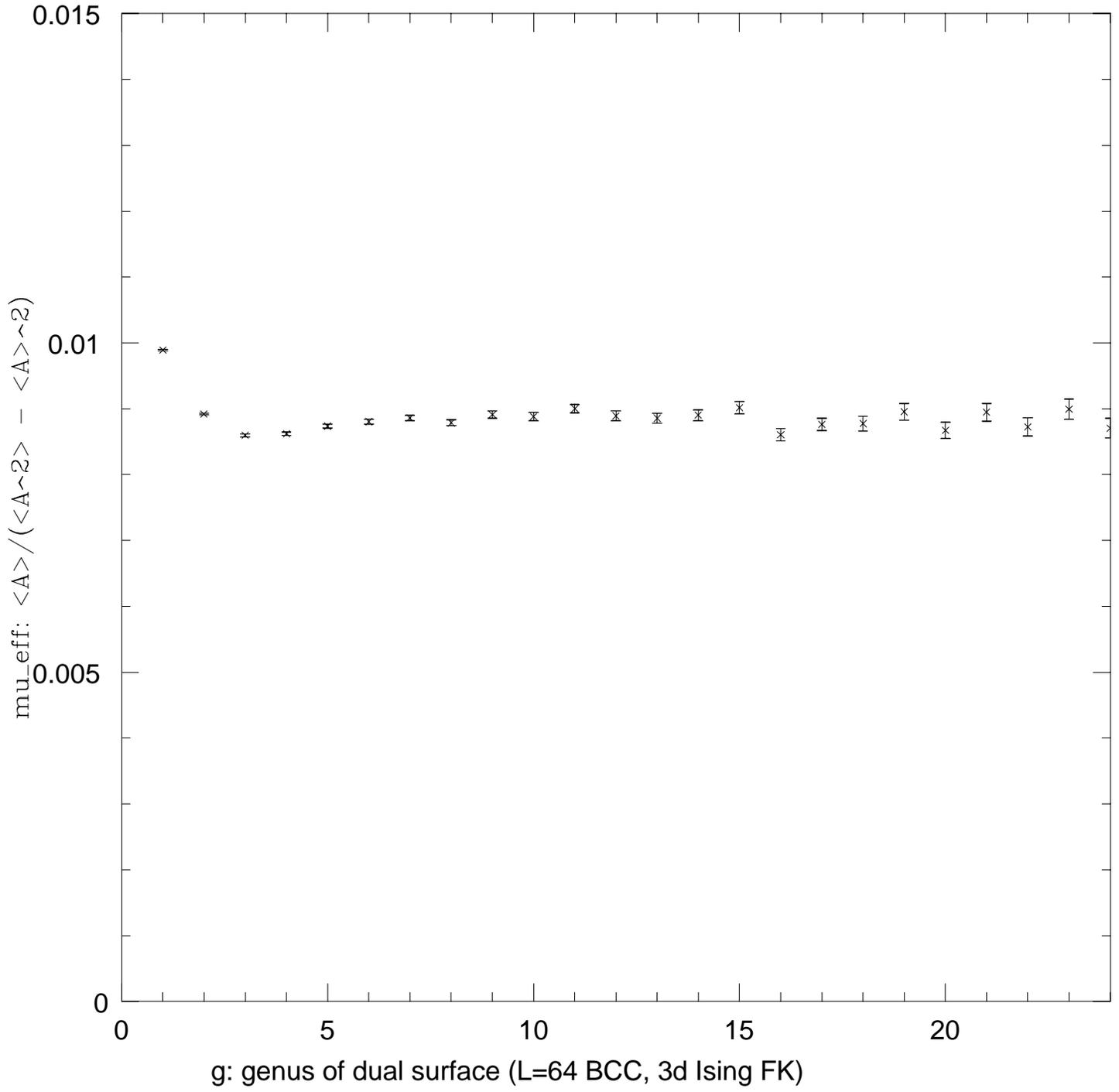
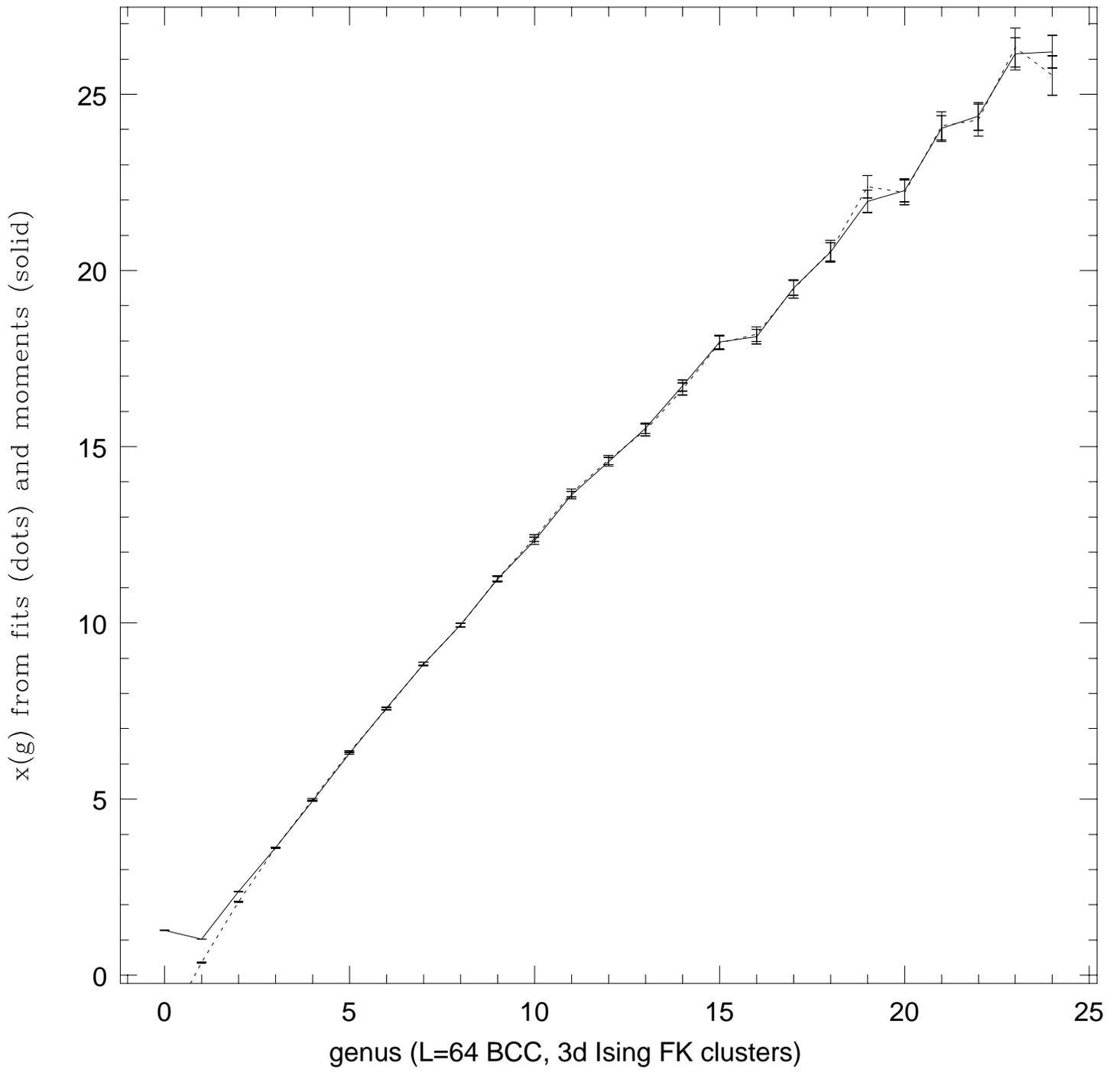
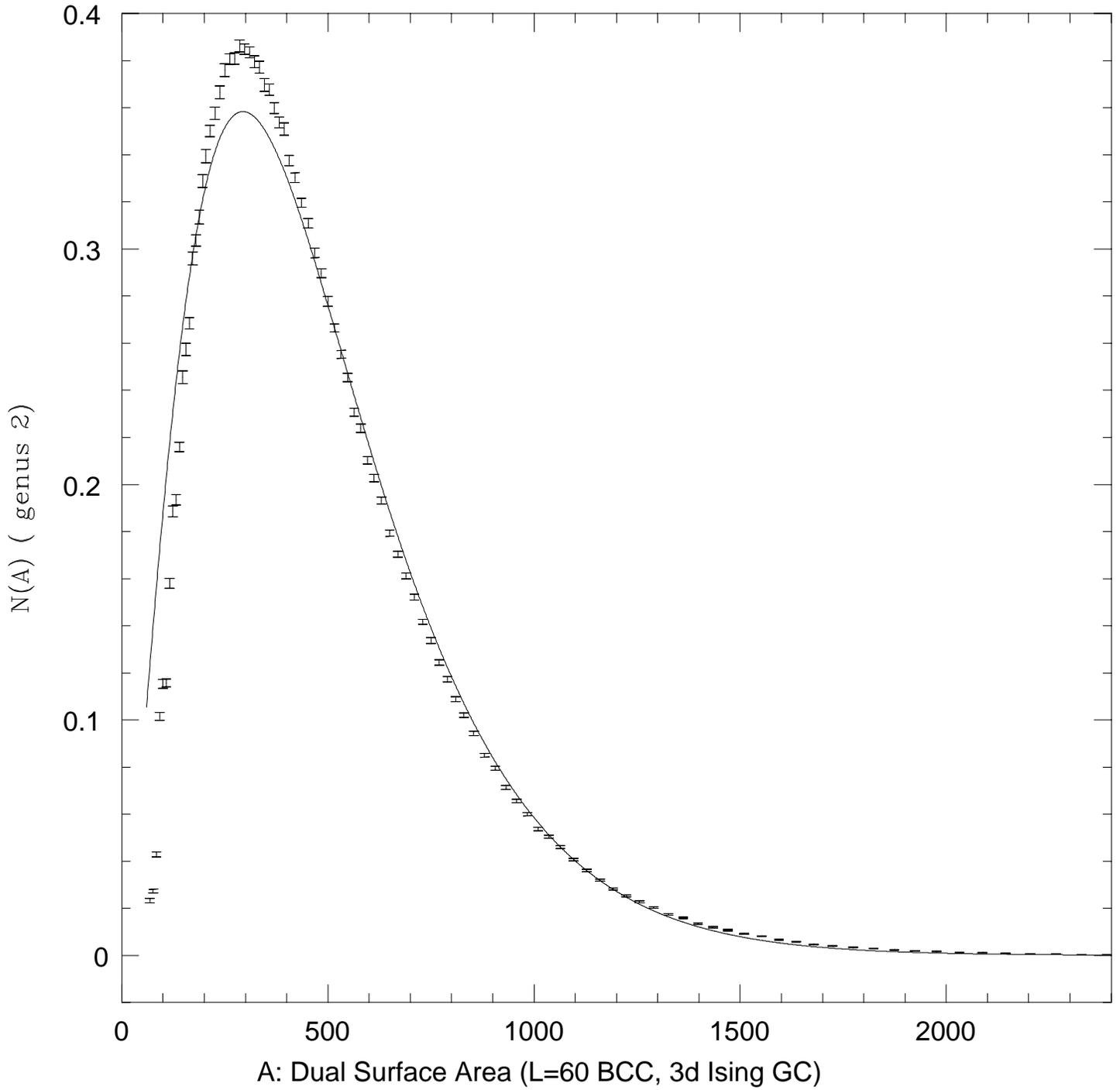
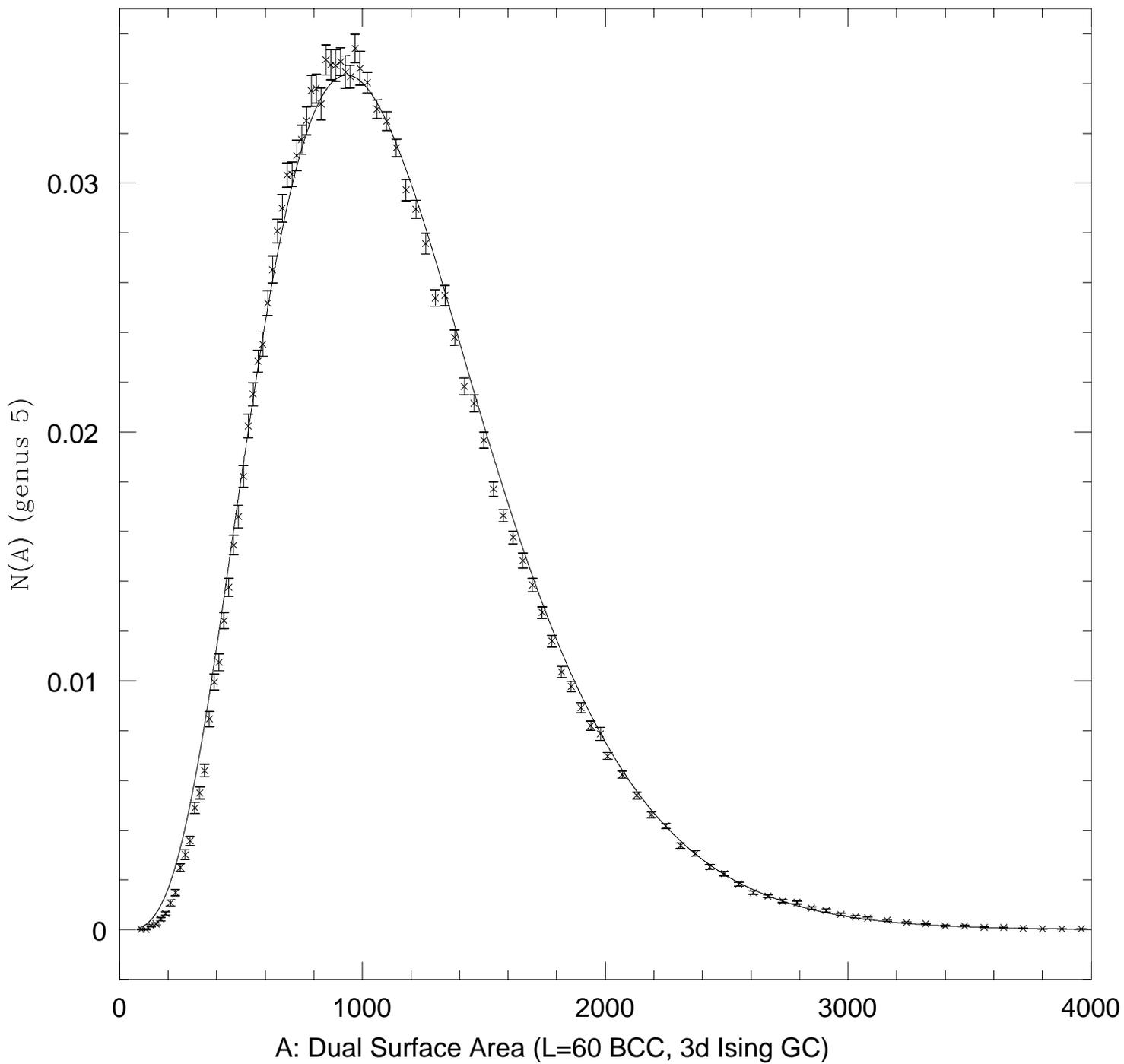
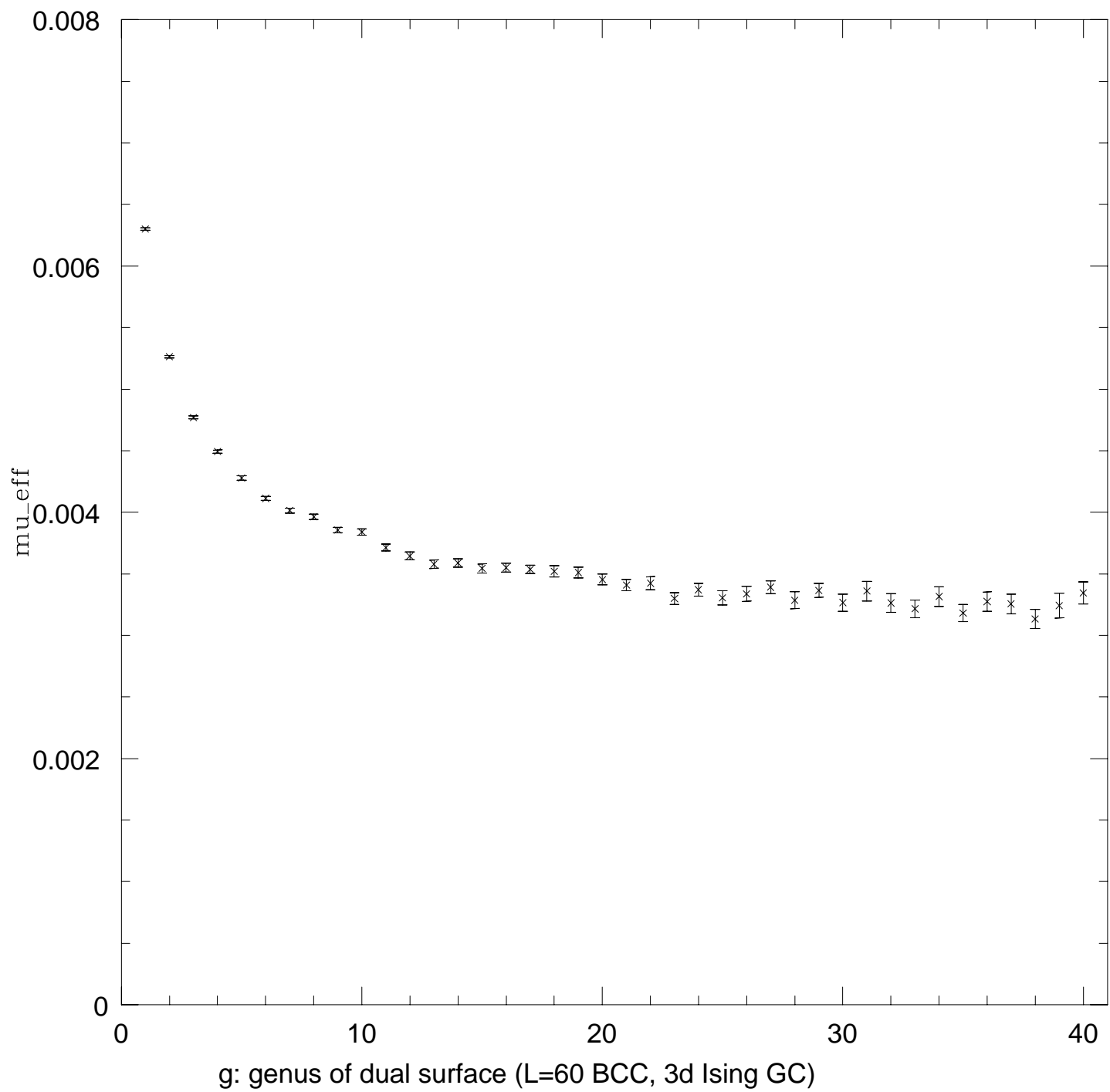
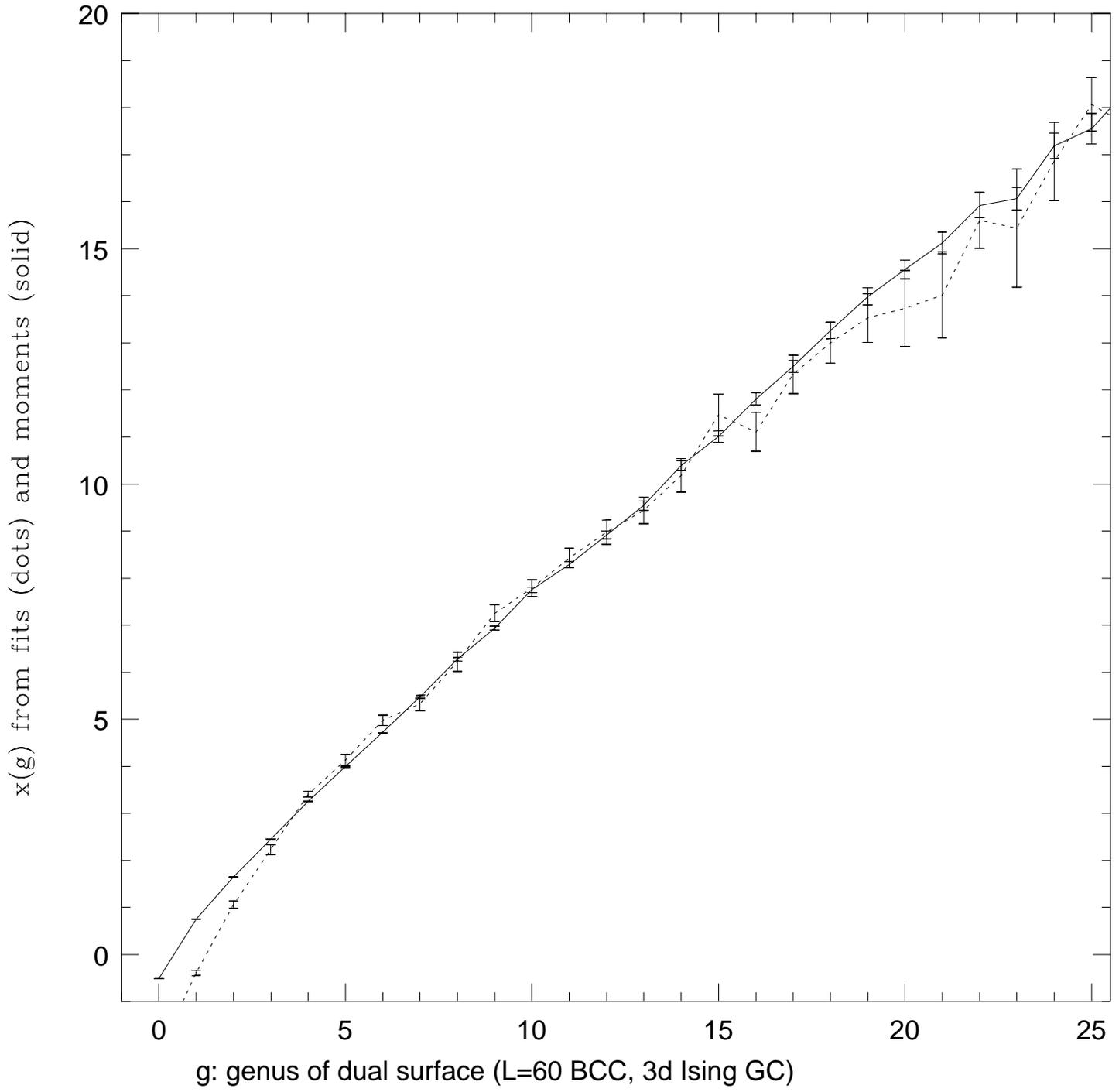
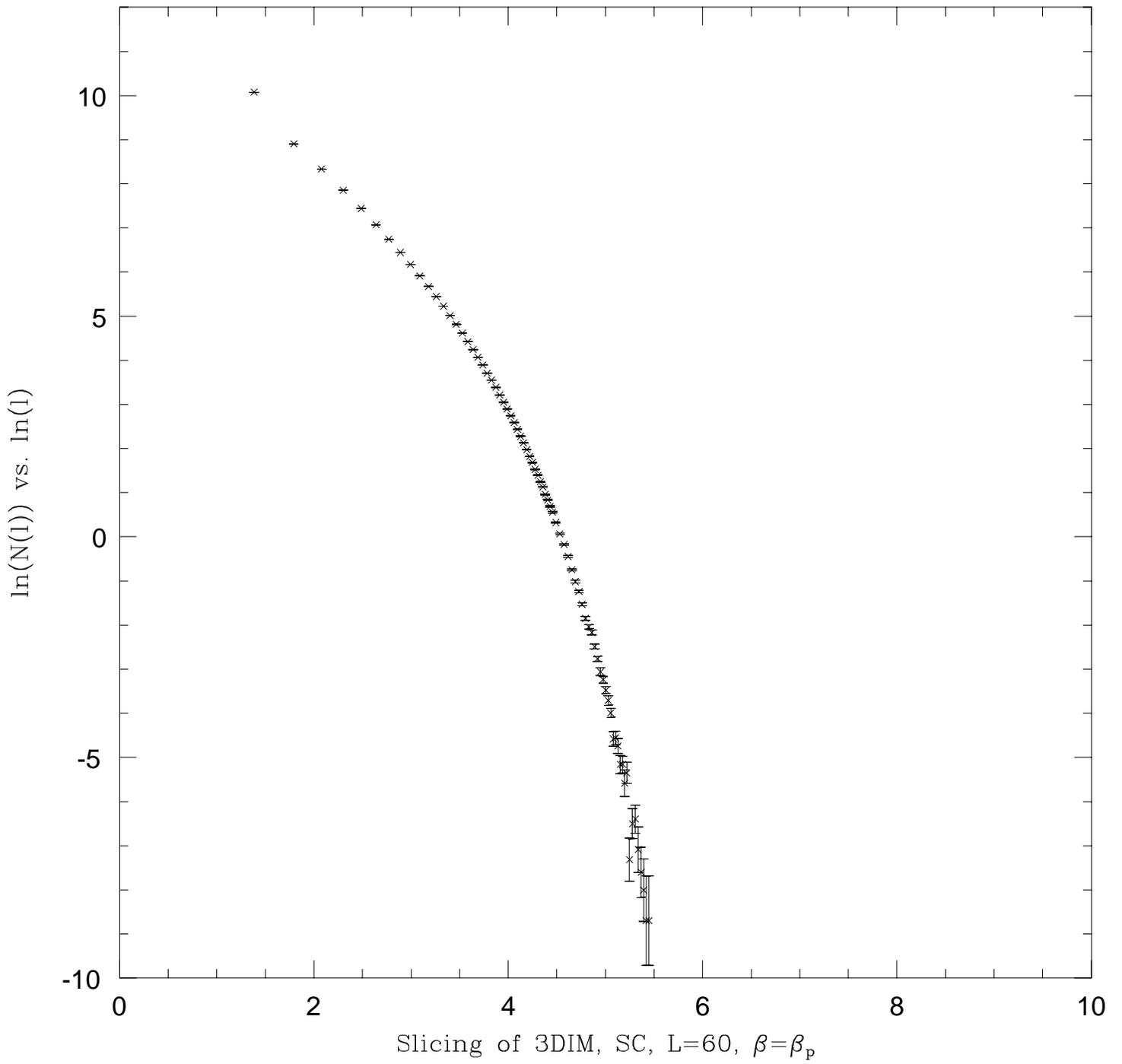
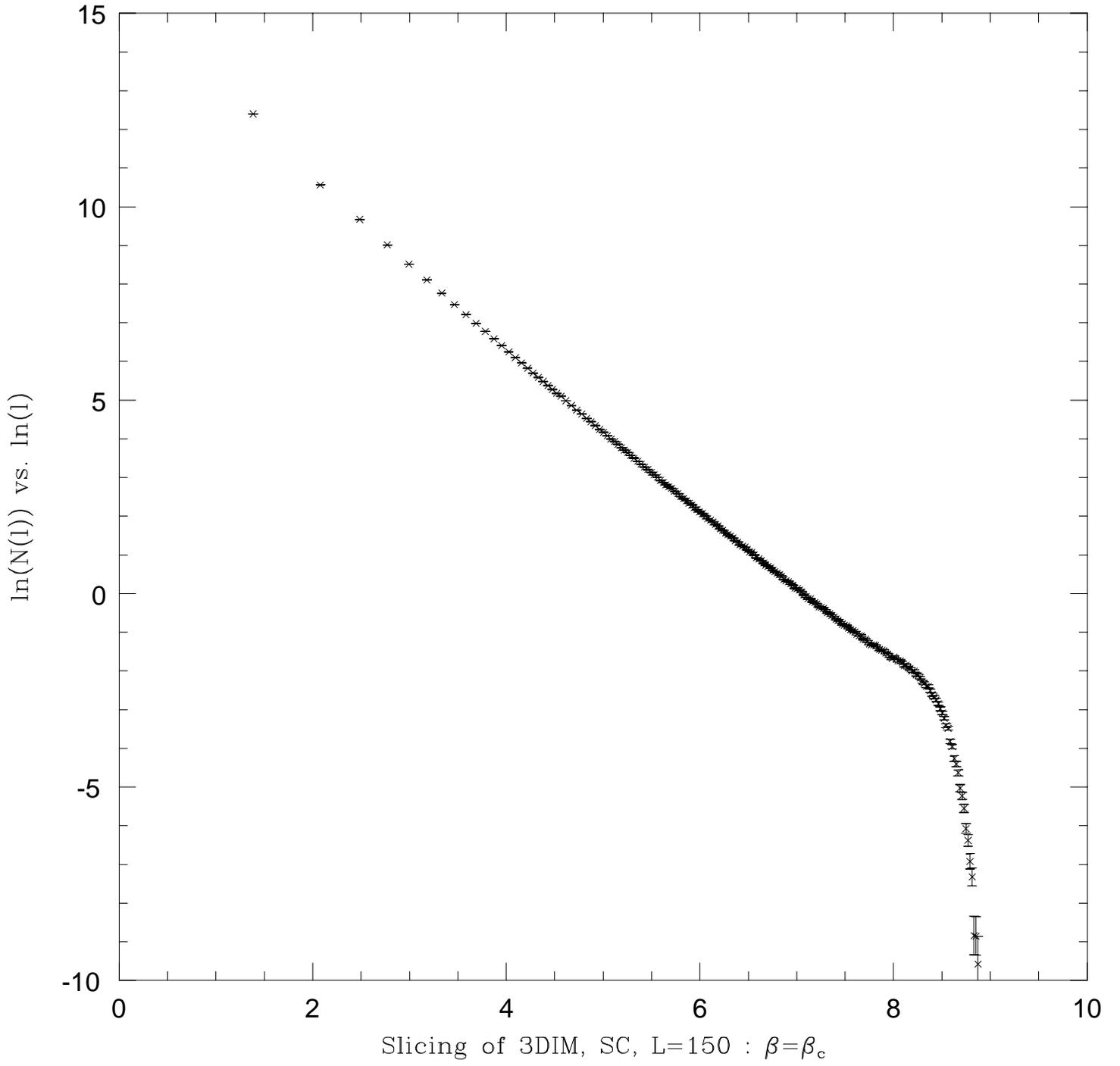
\end{document}